# Fiber-optic diagnostic system for future accelerator magnets

Maria Baldini[1*], Giorgio Ambrosio[1], Paolo Ferracin[3], Piyush Joshi[2], S. Krave[1], Linqing Luo[3], Maxim Marchevsky[3], G. Vallone[3], Xiaorong Wang[3]

[1] *Fermi National Accelerator laboratory, Batavia, IL 60510*
[2] *Brookhaven National Laboratory, Upton, NY 11973*
[3] *Lawrence Berkeley National Laboratory, Berkeley, CA 94720*
*mbaldini@fnal.gov

EXECUTIVE SUMMARY

The next generation high energy physics accelerators will require magnetic fields at ~20 T. HTS coils will be an essential component of future accelerator magnets and several efforts are currently dedicated on designing 20 T HTS- LTS hybrid magnets. Among the existing challenges, there is the lack of a robust quench detection system for hybrid magnet technology. Indeed, present quench detection systems based on growth of resistive voltage are not effective for HTS coils. Another big challenge is represented by the high number of training quenches required by $Nb_3Sn$ magnets to reach performance level.

In this framework it is important to find a tool that allow local real-time monitoring of magnet strain and temperature. In this paper, we propose the use of fiber optics sensors for diagnostic and quench detection in future accelerator superconducting magnets. Discrete and distributed fiber optic sensors have demonstrated to be a promising tool. However, if the goal is to instrument hundreds of accelerator superconducting magnets and to move beyond the proof-of-concept level, significant developments are still needed. Here, we are going to present the most recent results and discuss the most urgent technical developments in order to make those sensors a robust and reliable diagnostic tool for accelerator superconducting magnets over the next 10 year.

We foresee that discrete fiber sensors will be a stable diagnostic probe for superconducting magnets over the next 3 t0 5 years. More R&D work will be necessary for distributed fibers. Those sensors are extremely promising tools since they can be wound together with a coil and used to measure strain and temperature variations with a spatial resolution of around 0.5 mm. The most urgent needs are the increase of sample rate and sensitivity. Close collaboration with vendors will be necessary to improve mechanical properties and fabrication processes in order to produce hundreds of meters of fiber and instrument a large number of accelerator superconducting magnets. This R&D efforts will last up to 10 years with a founding level that spans between 5-10 M$.

## INTRODUCTION

Accelerator magnets fabricated using high temperature superconductors are going to be an essential element for the next generation of high energy physics machines (Muon Collider, HE-LHC, future proton-proton collider).

Indeed, Muon collider will require the employment of magnets capable to operate at a high ramping rate that cannot be guaranteed with the present NbTi and $Nb_3Sn$ technology. Future hadron-hadron colliders will need a magnetic field above 16 T. Indeed, one of the high priorities identified by the European Strategy for Particle physics has been the ramping up of the R&D efforts on developing high field superconducting magnet. A very similar goal has been identified by the US Magnet development program (US-MDP).

The NbTi dipole magnets currently installed in the LHC can generate a 9 T field. The US Accelerator Upgrade project together with CERN is fabricating 25 $Nb_3Sn$ quadrupoles for the upgrade of the LHC interaction region [1]. Those quadrupoles are the state of the art of $Nb_3Sn$ technology with a peak magnetic field of around 12 T. In within the US-MDP program, several efforts have been dedicated to the fabrication $Nb_3Sn$ dipole with a field > 15T. A 14.5 T record field has been recently measured during a vertical test in superfluid He carried on at Fermi National Laboratory [2]. In this scenario, it is evident that HTS magnets will be the fundamental ingredient to produce a field > 16 T for future HEP machines.

Considering the high cost of high temperature conductors: REBCO and Bi-2212, several R&D efforts have been already focused in designing hybrid magnets [3-7], with a field of 11-12 T generated by $Nb_3Sn$ dipole and a field of 5 T generated by the HTS magnet insert. Hybrid magnets represent the best opportunity of generating field above 20 T.

A magnetic field > 20 T presents several challenges. In this paper we will focus on two main issues: identify the factors affecting magnet performance in those challenging conditions in order to drive the design and reduce the number of quenches and the lack of a robust quench detection system for hybrid magnet technology.

With present $Nb_3Sn$ technology, magnet training requires up to 50 quenches. For a future FCC-like machine, this means a total number of training quenches above 200,000 resulting in a prohibitive impact on cost and schedule. Novel diagnostic approaches are necessary to obtain quantitative information on magnet quench mechanisms, disturbance spectrum and mechanical limits. For example, this information will allow to fine tune of magnet preload and coil-pole interface to significantly reduce the number of quenches.

Quench development and propagation are much slower in HTS coils: the order of magnitude being around mm/s for HTS [8,9] and around m/s for LTS magnets. Discrete voltage taps are ineffective for slow quench propagation and local temperature in HTS system may cross the safe threshold for conductor degradation before a quench is detected. It is then necessary to develop a novel detection system for HTS coils.

In both cases, local real-time monitoring of magnet strain and temperature is needed.

Over the past 20 years, fiber optic sensors have been identified as a very promising diagnostic tool for superconductive magnets [10]. Fiber optics are cheap and well-established technologies employed in several industrial sectors. The working principle of those sensors is very simple: the spectral shift observed in the fiber can be directly connected to strain and temperature variations [11]. Those sensors can be divided in two main categories: discrete sensors based on Bragg grating principle (FBG) or distributed sensors based on Rayleigh, Raman or Brilluoin backscattering. The employment of fibers presents several advantages: fibers for example are not sensitive to electromagnetic fields.

In this white paper, we propose the development of fiber optics diagnostic sensors for diagnostic and quench detection in future accelerator superconducting magnets. We are going to present some of the most recent results and discuss the ongoing issues and the technical developments required over the next 10 year in order to make those sensors a robust and reliable diagnostic tool for accelerator superconducting magnets.

# FIBER OPTICS SENSOR FOR SUPERCONDUCTING MAGNET DIAGNOSTIC

Several studies have demonstrated the potential use of fiber optic sensors in $Nb_3Sn$ magnets. FBG (fiber Bragg gratings) fibers are currently used at CERN to measure coil and structure strain in the IR low-$\beta$ focusing $Nb_3Sn$ quadrupole under development for the HL-LHC upgrade [12]. Those sensors are used to monitor strain variations over the entire lifetime of a magnet: cooldown, energization and for the first time during a quench since they are insensitive to electromagnetic forces [12, 13].

The vertical magnet test facility at FNAL has been recently modified to accommodate up to 16 fiber sensors (see Fig. 1). The first cold test at 1.9 K has been performed in March 2022 on a mirror magnet.

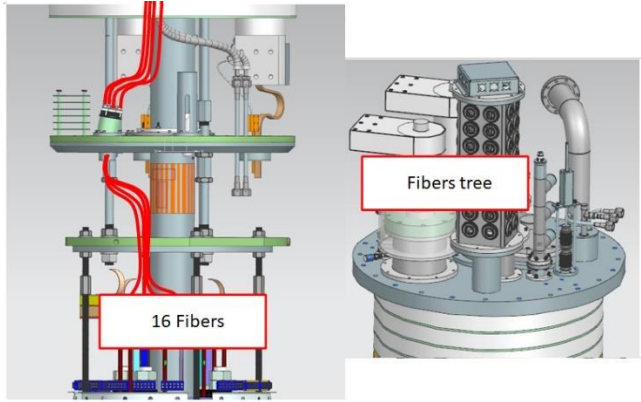

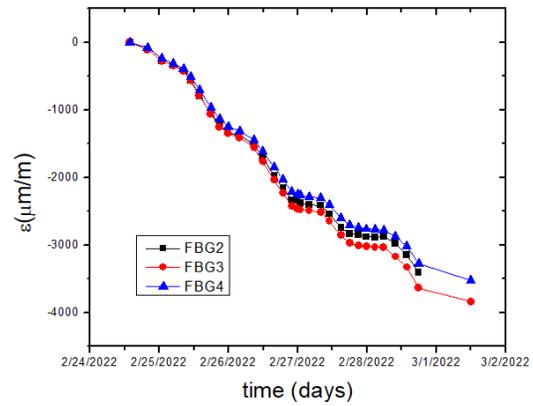

*Figure 1: modifications of the Vertical test facility to accommodate fiber optics sensors: fiber G10 plug at the level of the lambda plate and a new tree has been designed to extract fiber sensor signals from the magnet during testing.*

*Figure 2: Absolute strain variation during cooldown collected from three different FBG sensors.*

The data were collected using a fiber with four FBG sensors installed on the stainless-steel shell. In Fig. 2, the strain data collected during cooldown are reported.

The Nb3Sn MQXFA superconducting magnets fabricated through the HL-LHC Accelerator Upgrade project (AUP) are currently being instrumented with FBG fibers. On top of measuring strain during magnet preload and training, those sensors will also allow to get information about strain on the coil during the LMQXFA coldmass shell welding. Unlike strain gauges, optical fibers are extremely thin (<200 um diameters) so they can survive the insertion of the magnet bore and allow to assess coil prestress increase after each welding pass. This capability is going to reduce significantly the risk of coil damage during welding.

Exploiting the synergies among the HL-LHC Accelerator Upgrade project (AUP), the Magnet Development Program (MDP) and Lab Directed R&D (LDRD) funds, fiber technology is being currently implemented across several US National Laboratories. We expect for FBG discrete sensors to become a stable diagnostic probe for superconducting magnets over the next 3 to 5 years.

Distributed fibers appear to be even a more promising diagnostic tool for strain detection in superconducting magnets. In Fig. 3, the strain variation of selected points of 2 m fiber glued on a steel bar are displayed. The data were collected during the application and the release of a 50 MPa tension. Data were taken with the highest spatial resolution (around half a millimeter). Strain can be collected on more than 100000 data point along a 10 m fiber. This capability is extremely valuable to identify the weaknesses of novel stress management structures as the CCT and SMCT coils that are currently under development in within the MDP program.

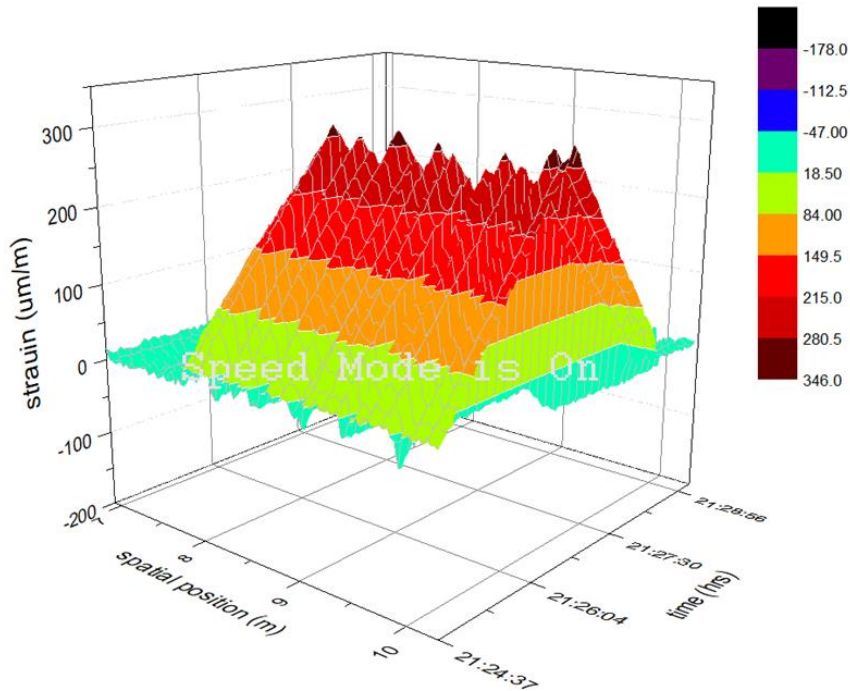

*Figure 3: Strain variation measured on selected spatial pitches of a 2 m fibers.*

Optical fibers can be used to measure both strain and temperature variations. Those sensors have been proved to be successfully wounded in a coil and to survive epoxy impregnation process [14]. Rayleigh fibers used as temperature sensors can have a two-fold aim; to measure the energy spectrum and to identify the exact quench location and the mechanical disturbances at the level wounded cable. The employment of a continuous strain/temperature sensor can help minimizing the required operating margin and significantly reducing training for superconducting magnets.

FIBER OPTICS SENSOR FOR QUENCH DETECTION
The greatest potential of Rayleigh sensors relays on the possibility of using them for quench detection in HTS magnets. Indeed, those sensors could provide significant advantages over traditional techniques for detecting normal zones in HTS [15]. Fibers were integrated into a REBCO conductor architecture and demonstrated strain sensing capabilities as well as thermal perturbation detection and localization with higher spatial resolution than taps [16]. Although those sensors have strong potential, recent results suggest that the claim of a shorter time response of optical fibers [15] compared to other diagnostic tools deserves further investigations [17].

We use Rayleigh distributed fiber optic sensing system to identify the locations of the resistive transition in a CORC® wire [18]. The optical fiber was co-wound with the sample, followed by epoxy impregnation (see Fig.4). During the test, the heating in the bent region of the

CORC® wire was successfully localized at a power level of 0.2 W which correspond to a 5kA and 40 uV voltage rise, a condition observed in recent REBCO magnet tests [19, 20].

In conclusion, optical fiber sensors are demonstrating to have a great potential for diagnostic and quench detection in superconducting magnets. Significant technical issues still need to be addressed to make this technology mature and robust to be integrated in future HEP machines. In the next session, we list what we believe are the three most urgent challenges to be addressed over the next 10 years.

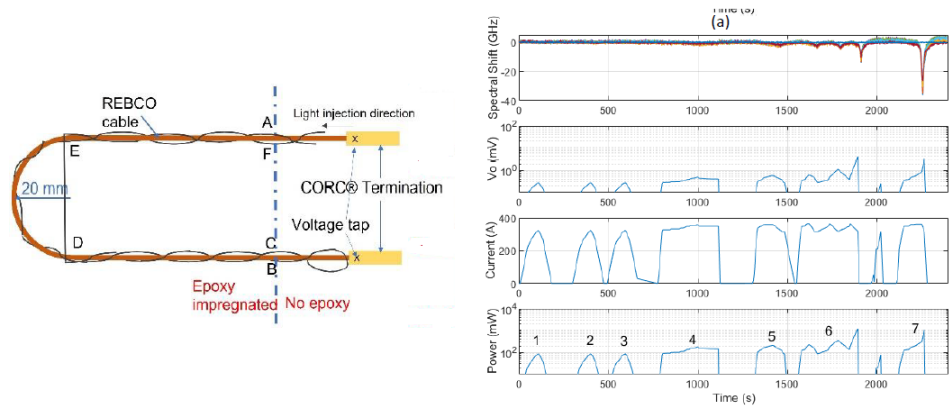

Fig. 4 Left: drawing of the CORC® wire and the fiber. Right: spectral shift observed at point E on the fiber. voltage across the CORC® wire; current and the total power generated [18].

## ONGOING ISSUES and DISCUSSION

IMPROVE SENSITIVITY: It is necessary to increase the sensitivity of those sensors. Indeed, results reported in ref[18] show that signals were too weak to be detected with no impregnation and with low power. Moreover, fiber optics have very low sensitivity to temperature below 50 K and future machines will probably work at relative low temperature (T < 4.5 K). The future R&D efforts should build on the work already performed on fiber coating for FBG sensors [21, 22, 23]. Small Business Innovation Research (SBIR) funds have been already awarded to some of the major players among fiber vendors. However, several more fast turn-around tests and collaboration with multiple optical fiber vendors are required to identify the best coating material to improve sensitivity and optimize signal and use those fibers as temperature sensors for quench detection.

A second issue is connected to the fact that strain and temperature effects cannot be separated in an easy way. Proof of principle experiment have demonstrated that the fiber can be wound and used in a strain free configuration [12]. For example, this can be done embedding the fiber in a Teflon tube. More robust and reproducible solutions need to be found to distinguish the two effects and to make sure those sensors can separately detect small temperature and strain variations.

IMPROVE RESOLUTION: A trade-off between spatial and temporal resolution needs to be identified to use Rayleigh sensors efficiently. At moment, distributed fibers can be as long as 50 m. However, the length of the fiber affects the sample rate. One of the most urgent need is to have long fibers (>100 m) with the highest sample rate (>1kHz) and highest spatial resolution (pitch <

1mm). This is a necessary step if we want to use those sensors for diagnostic or quench detection during a quench. Indeed, the capacity of detecting a quench is correlated to the real time signal processing of variations of the spectral shift. Technical solutions for improving the data acquisition system need to be found. The goal is to optimize the signal processing and increase the sample rate.

In this framework is also crucial to establish close collaboratiosn with the vendors that are currently the major players in developing fiber sensors for superconducting magnets such as LUNA Innovations and HBK FiberSensing.

INSTALLATION and SCALING UP: Fibers are fragile and can be fabricated with several coating materials. If any novel coating material with optimal sensitivity is going to be found, we need to be able to fabricate hundreds of meters of coated fibers. Mechanical properties, fabrication and installation processes need to be scaled up to industrial level in order to instrument a large number of accelerator superconducting magnets.

CONCLUSIONS

At present fibers have proven to be a promising tool at the level of proof of principle experiments. Over the next 3-5 years, we expect for FBG fiber optic sensors to become a robust and standard diagnostic tool for measuring strain on the shell and on the coil pole of superconducting magnets. More R&D development will be necessary for employment of Rayleigh distributed sensors. The work performed over the next 10 years will be crucial to identify optimal materials to improve fiber sensitivity and to increase spatial resolution, and sample rate. Novel technical solutions need to be found to make those sensors a robust and reliable tool.

The R&D magnets fabricated in within the US-MDP program and future programs provide excellent testbeds for the proposed studies. Hybrid magnets prototype will be fabricated over the next 5-10 years. Spatial/temporal resolution of distributed optical sensors can be tuned and modified according to the needs, making those fibers the most promising quench detection system for LTS-HTS hybrid magnets where integration is the major challenge. The long-term goal would be to integrate fibers in above 20 T hybrid magnets making them a solid quench detection system for future HEP machines.